\documentclass[12pt]{article}

\usepackage{amsmath,amssymb,cite}

\global\arraycolsep=1pt
\numberwithin{equation}{section}
\newcommand{\bel}[1]{\begin{equation}\label{#1}}                     
\newcommand{\bal}[1]{\begin{eqnarray}\label{#1}}                     
\newcommand{\be}{\begin{equation}}
\newcommand{\ee}{\end{equation}}

\newcommand{\im}{\mathrm{i}}
\newcommand{\ex}{\mathrm{e}}
\newcommand{\de}{\mathrm{d}}

\newcommand{\qq}{\qquad}
\newcommand{\mat}[1]{\begin{pmatrix} #1 \end{pmatrix}}
\renewcommand{\thefootnote}{\fnsymbol{footnote}}

\begin{document}

%
%
\begin{titlepage}

\begin{flushright}
       \normalsize
       November, 2004  \\

\end{flushright}

\begin{center}
{\Large \bf  Comments on Equations of Motion for Pure Spinors in
Even Dimensions}
\end{center}
\vfill
\begin{center}
{
{Takeshi Oota${}^1$}\footnote{
e-mail address: {\tt toota@sci.osaka-cu.ac.jp}}\footnote{
Address after 7 November: \\
Osaka City University, 
Advanced Mathematical Institute, \\
3-3-138 Sugimoto, Osaka 555-8585, Japan }
        }\\
\end{center}

\vfill
\begin{center}
     ${}^1$\it Istituto Nazionale di Fisica Nucleare,
       Sezione di Bologna, \\
       Via Irnerio 46, 40126 Bologna, Italy
~\\
\end{center}

\vfill


\begin{abstract}
A Berkovits type action for pure spinors in even dimensions
is considered. The equations of motion for pure spinors are
investigated by using explicit parameterizations which solve
the pure spinor constraints. 

For general interactions, the equations of motions are shown 
to be modified from the naive ones. The extra terms contain
a particular projector $\Delta(\lambda, \omega)$.

If the interactions are restricted to the ``ghost number'' $u(1)$
and the Lorentz $so(p,q)$ current couplings, the action has a large
``gauge symmetry''. In this case, in some ``gauges'',
the extra terms vanish and
the equations of motion for the pure spinors retain 
the naive form even if the pure spinor constraints are taken into account.

\end{abstract}

\vfill

\end{titlepage}

\renewcommand{\thefootnote}{\arabic{footnote}}
\setcounter{footnote}{0}


\section{Introduction}

Last year, hidden infinite-dimensional symmetries were found 
\cite{BPR}
in the classical Green-Schwarz superstring 
on the $AdS_5 \times S^5$ background \cite{MT}. 
Some properties of the non-local classical conserved charges
are investigated in the pp-wave limit \cite{ald}.
And recently, it is shown that the hidden symmetries 
are classical super Yangian \cite{hat-yos}.

An important problem is to check that this classical
integrability survives the quantization or not. 
But due to the $\kappa$-symmetry, it is difficult
to quantize the Green-Schwarz action in 
the curved back ground with a Ramond-Ramond flux.

One promising approach to quantize superstring action is
the Berkovits' pure spinor formalism 
\cite{B00a,B00b,B00c,B02}\footnote{
Some related works,
such as extension of pure spinor formalism, 
can be found in 
\cite{AK1,AK2,AK3,OT,MMOST,GPPvN,GPvN02a,GPvN02b,GPvN03a}.}.
The existence of the hidden
integrability is shown also for this formalism \cite{vallilo03}.

In the pure spinor formalism, nilpotency constraints for
the BRST-like charges and the holomorphicity constraints
for the associated currents play crucial role in
restricting the geometry of the target space \cite{ber-how}.
The equivalence of these
constraints to the on-shell supergravity constraints is analyzed.

Before going to consider the problem of quantum integrability
in the particular superstring background,
we would like to point out that
there is a subtlety in the holomorphicity constraints,
which comes from the pure spinor constraints 
$\lambda^{\alpha} ( \gamma^a)_{\alpha \beta} \lambda^{\beta} = 0$.

In the superstring action of the pure spinor formalism,
the pure spinor $\lambda^{\alpha}$ and its pair $\omega_{\alpha}$
appear in the following form (in the conformal gauge): 
\bel{Ipure}
I = - \frac{1}{\pi \alpha'}
\int \de^2 \xi \left(
\omega_{\alpha } \overline{\partial} \lambda^{\alpha}
+ \omega_{\alpha } X^{\alpha}{}_{\beta} \lambda^{\beta}
\right).
\ee
Here $X^{\alpha}{}_{\beta}$ is a field which takes
values in the Clifford algebra with the even degree, and is
independent of $\lambda^{\alpha}$
and $\omega_{\alpha}$. 

In derivation of the holomorphicity constraints \cite{ber-how}, 
the naive equations of motion
\be
\overline{\partial} \lambda^{\alpha} + X^{\alpha}{}_{\beta}
\lambda^{\beta} = 0,
\ee
\be
- \overline{\partial} \omega_{\beta}
+ \omega_{\alpha} X^{\alpha}{}_{\beta} 
= 0,
\ee
are used.\footnote{Strictly speaking, 
the equations of motion for pure spinors themselves are not necessary.
The equations of motion for the ghost number current and
those for the Lorentz currents are sufficient 
for deriving the holomorphicity constraints \cite{Bc}.}

But, it is quite common that for a system with constraints,
the equations of motion are different from the naive ones.
Let us recall the case of the $O(N)$ invariant
non-linear sigma model : $\mathcal{L}_{O(N)}
= (1/2)\partial_{\mu} \phi^I \partial^{\mu} \phi^I$.
$N$ real scalar fields $\phi^I$ $(I=1,2,\dots, N)$ are
required to obey the constraint $\phi^I \phi^I = 1$.
With help of the Lagrange multiplier, 
the equations of motion are obtained as
$\partial_{\mu} \partial^{\mu} \phi^I -  
\phi^I (\phi^J \partial_{\nu} \partial^{\nu} \phi^J) = 0$.
The second term is the effect of the constraint.
Or, equivalently, we can derive equations of motion by solving 
the constraint. If we use a solution $\phi^N = \pm 
\sqrt{1 -\phi^i \phi^i}$ and substituting it into $\mathcal{L}_{O(N)}$,
the Lagrangian for $N-1$ real 
scalars $\phi^i$ ($i=1,2,\dots, N-1$) is obtained:
$\mathcal{L}_{O(N)} = (1/2) G_{ij}(\phi) \partial_{\mu} \phi^i
\partial^{\nu} \phi^j$, where $G_{ij}(\phi)
= \delta_{ij} + \phi^i \phi^j/(1 - \phi^k \phi^k)$.
By using this Lagrangian, we get the equations of motion:
$\partial_{\mu} \partial^{\mu} \phi^i
+ \phi^i G_{jk} \partial_{\mu} \phi^j \partial^{\mu} \phi^k = 0$
which are equivalent to the manifestly $O(N)$ invariant ones.

Therefore, it seems important to consider the
effect of pure spinor constraints to the equations of motion.
In order to make the meaning of the effect from the 
pure spinor constraints clear,
we consider a pure spinor action of the type \eqref{Ipure}
in general even dimension ($D=2N$).

The approach we take in this paper is an analog of the second one
for the case of $O(N)$ invariant non-linear sigma model.

In section $2$, we explain our choice of the pure spinor action
which has ``gauge symmetries''.
The explicit parametrization for $\lambda$ and $\omega$
are introduced. Transformation properties of these
parameters for the finite Lorentz transformation $SO(p,q)$
are summarized.

Because it is rather difficult to treat the 
pure spinor constraints by the Lagrange multiplier method,
we use an explicit parametrization of 
pure spinors which solves the constraints.
Using the explicit local coordinates
$(\lambda, u_{rs})$ and $(\beta, v^{rs})$
for a pure spinor $\lambda$ and its pair $\omega$, 
the equations of motion for the pure spinors are derived
in the simple ``bosonic ghost'' gauge.

The case for the action without the gauge symmetry is discussed in
section $3$. In this case, $\omega$ is also a pure spinor.
The equations of motion for pure spinor pairs $(\lambda, \omega)$
are derived for the pure ``bosonic ghost'' parametrization.
 
Section $4$ is devoted to discussions.

In Appendix A, some useful properties of pure spinors
in even dimensions are summarized.

\section{Pure Spinor Action and Equations of motion}

\subsection{Pure spinor action with a ``gauge symmetry''}

In this section, we consider a pure spinor action 
of the Berkovits type \eqref{Ipure}
for a target space with even dimension ($D=2N$). 
Let us consider the following Lagrangian for a (bosonic) pure spinor
$\lambda$ and its (bosonic) pair spinor $\omega$
(in the conformal gauge):
\be
\mathcal{L} = \langle \omega^{\mathrm{T}} \mathcal{C}| 
\left( \overline{\partial} + X \right) | \lambda \rangle.
\ee
Here $\langle \omega^{\mathrm{T}} \mathcal{C}|
= ( | \omega \rangle)^{\mathrm{T}} \mathcal{C}$, 
$\mathcal{C}$ is the charge conjugation matrix :
$( \Gamma^a)^T = \mathcal{C} \Gamma^a \mathcal{C}^{-1}$ 
($a=1,2,\dotsc, 2N$) and $X$ is a field which takes
values in the Clifford algebra with even degree.

We choose the chirality of the pure spinor $|\lambda \rangle$
to be positive. Then, the chirality of the spinor $|\omega \rangle$
equals to $(-1)^N$.

We consider the pure spinor $\lambda$ 
``near'' the Fock vacuum $| + \rangle$ and
take the local coordinate $(\gamma, u_{rs})$ as follows
(for details, see appendix A):
\bel{purelambda}
| \lambda \rangle = \gamma \ex^U | + \rangle, \qq
U = \frac{1}{2} u_{rs} A^{(-)}_r A^{(-)}_s.
\ee
We choose $|- \rangle := A_1^{(-)} A_2^{(-)} \dotsm A_N^{(-)}
| + \rangle$.
Let us denote the Hermitian conjugated and transposed states of 
$| \pm \rangle$ by
$\langle \pm| = ( | \pm \rangle)^{\dag}$
and $\langle \pm, \mathrm{T} |= ( | \pm \rangle)^{\mathrm{T}}$
respectively.
The normalization of the charge conjugation matrix $\mathcal{C}$
is fixed by the condition:
$\langle -, \mathrm{T} | \mathcal{C}
= \langle +|$. Then, $\langle +, \mathrm{T} | \mathcal{C} = \varepsilon_N
\langle - |$. Here $\varepsilon_N:=(-1)^{(1/2)N(N-1)}$.

Note that $U$ is a linear combination (with coefficients in $\mathbb{C}$) of
the Lorentz generators in the spinor representation: $(1/2) \Gamma^{ab}$.
So, it may be possible to  interpret the relation 
$\ex^{-U} \Gamma^a \ex^U = \Gamma^b \Lambda_b{}^a(u)$
as a ``complexified'' Lorentz transformation.

We can easily see that
\be
\langle \lambda^{\mathrm{T}} \mathcal{C} |
\Gamma^{a_1 a_2 \dotsm a_n} | \lambda \rangle
= \varepsilon_N  \gamma^2
\langle - | \ex^{-U} \Gamma^{a_1 a_2 \dots a_n} \ex^U | + \rangle 
= 0, \qq \text{for } \ \ 0 \leq n \leq N-1.
\ee
Similarly,
\be
\langle \lambda^{\mathrm{T}} \mathcal{C} | 
\Gamma^{a_1 a_2 \dotsm a_n} \overline{\partial}
| \lambda \rangle = 0, \qq \text{for } \ \ 0 \leq n \leq N-3.
\ee
Therefore, the kinetic part $\mathcal{L}_{\mathrm{kin}}
= \langle \omega^{\mathrm{T}} \mathcal{C}
| \overline{\partial} | \lambda \rangle$
is invariant under ``gauge transformations'':
\bel{GTr}
| \omega \rangle \rightarrow | \omega' \rangle
+ \xi | \lambda \rangle,
\ee
where
\be
\xi = \sum_{\substack{ 0 \leq n \leq N-3 \\ N-n = \text{even}} }
\frac{1}{n!} \xi_{a_1 a_2 \dotsm a_n}
\Gamma^{a_1 a_2 \dots a_n}.
\ee
The restriction on $N-n$ comes from the
chirality condition:
$\overline{\Gamma} | \omega' \rangle = (-1)^N | \omega' \rangle$.

As in the pure spinor action \eqref{Ipure},
we require that the rest part of the Lagrangian also
respects this ``gauge symmetry''. In other words, we restrict
$X$ as follows:
\bel{rX}
X = m + \frac{1}{2} f_{ab} \Gamma^{ab}.
\ee
It means that the pure spinor coupling to external sources
is allowed only through ``ghost number'' $U(1)$ 
current $j= \langle \omega^T \mathcal{C} | \lambda \rangle$
and $so(p,q)$ current $M^{ab} = \langle \omega^T \mathcal{C} |
\Gamma^{ab} | \lambda \rangle$.

Then, due to the gauge symmetry, 
the ``physical'' degree of the freedom of $|\omega \rangle$
becomes
$(1/2)N(N-1)+1$ and is equal to that of $|\lambda \rangle$.

Let us denote the components of $|\omega \rangle$ by
\be
| \omega \rangle = \omega_- | - \rangle + 
\frac{1}{2} \omega^{rs} A^{(+)}_r A^{(+)}_s | - \rangle
+ \frac{1}{4} \omega^{pqrs} A^{(+)}_p A^{(+)}_q
A^{(+)}_r A^{(+)}_s | - \rangle + \dotsm.
\ee
There are various ways to fix the gauge symmetry.
One choice is to set $|\omega \rangle$ to be a pure spinor:
\be
\left( A_r^{(-)} + A_s^{(+)} w_{sr} \right) | \omega \rangle = 0.
\ee
In the pure spinor gauge, we can take 
local coordinates $(\rho, w_{rs})$ for $|\omega \rangle$:
\bel{pureomega}
| \omega \rangle_{\mathrm{p}} 
= \rho \, \ex^W | - \rangle, \qq
W = \frac{1}{2} w_{rs} A_r^{(+)} A_s^{(+)}.
\ee

Other simple gauge is obtained by
setting as many components to be zero as possible:
\bel{simple}
| \omega \rangle_{\mathrm{s}} = \omega_- | - \rangle
+ \frac{1}{2} \omega^{rs} A_r^{(+)} A_s^{(+)} | - \rangle.
\ee

The components in these two gauge are related as follows:
\be
\omega_-= \rho, \qq \omega^{rs} = \rho w_{rs}.
\ee

For a fixed $(\gamma, u_{rs})$ \eqref{purelambda}, 
there are special ways to parametrize $\omega$ such that the
kinetic term $\mathcal{L}_{\mathrm{kin}}$ becomes 
that for a  collection of
bosonic ``$\beta \gamma$-ghosts'': 
\be
\mathcal{L}_{\text{kin}} = \beta \overline{\partial} \gamma 
+ \frac{1}{2} v^{rs} \overline{\partial} u_{rs}
= \beta \overline{\partial} \gamma 
- \frac{1}{2} \mathrm{tr}( v \overline{\partial} u).
\ee
In the superstring theories of the pure spinor formalism,
this ``bosonic ghost'' gauge plays crucial role
in quantization.

Within the pure spinor gauge \eqref{pureomega}, 
the pure ``bosonic ghost'' parametrization is given by
\bel{pbg}
| \omega \rangle_{\mathrm{p.b}} = \beta \ex^U \ex^Y | - \rangle
= \rho \ex^W | - \rangle,
\ee
where
\be
\rho = \beta \, \mathrm{det}^{1/2}( 1_N + uy),
\ee
\be
Y = \frac{1}{2} y^{rs} A^{(+)}_r A^{(+)}_s, \qq
y^{rs} = ( \beta \gamma)^{-1} v^{rs}, \qq
w = y ( 1_N + uy )^{-1} 
= v \bigl( \beta \gamma \, 1_N + u v \bigr)^{-1}.
\ee

Within the simple gauge \eqref{simple}, 
the simple ``bosonic ghost'' parametrization is given by
\bel{omegaghost}
\omega_- = \beta - \frac{1}{2} \gamma^{-1} v^{rs} u_{rs}, \qq
\omega^{rs} = \gamma^{-1} v^{rs}.
\ee
This corresponds to the following parametrization of $\omega$:
\be
{}_{\mathrm{s.b.}}\langle \omega^{\mathrm{T}} \mathcal{C} |
= \langle + | ( \beta - \gamma^{-1} V) \ex^{-U},\qq
V:= \frac{1}{2} v^{rs} A^{(+)}_r A^{(+)}_s.
\ee

\subsection{Lorentz transformation properties for
various parameters}

Before deriving equations of motion, we would like to
summarize the Lorentz transformation properties of
local coordinates.

Under the Lorentz transformation $\Lambda \in SO(p,q)$,
the isotropic complex vectors transform linearly
(see appendix):
$n'^{(I)}_{a} = \Lambda_{a}{}^{b} n^{(I)}_{b}$.
Let $\widehat{\Lambda}_{\mu}{}^{\nu}:= \omega_{\mu} \Lambda_{\mu}{}^{\nu}
\omega_{\nu}^{-1}$ (no sum), and
$a_{rs} := \widehat{\Lambda}_{2r-1}{}^{2s-1}$,
$b_{rs} := \widehat{\Lambda}_{2r-1}{}^{2s}$, 
$c_{rs} := \widehat{\Lambda}_{2r}{}^{2s-1}$ and
$d_{rs} := \widehat{\Lambda}_{2r}{}^{2s}$.
The condition $\Lambda \in SO(p,q)$ is converted into
$a^{\mathrm{T}} a + c^{\mathrm{T}} c = 1_N$, 
$b^{\mathrm{T}} b + d^{\mathrm{T}} d = 1_N$,
$a^{\mathrm{T}} b + c^{\mathrm{T}} d = 0$.

In the complex basis, the Lorentz transformation of the
isotropic vectors can be written as 
\be
\mat{ U' \\ T' } 
=  M(\Lambda) \mat{ U \\ T } 
= \mat{ A(\Lambda) & \ & B(\Lambda) \\
C(\Lambda) & \ & D(\Lambda) } \mat{ U \\ T },
\ee
where
\be
\begin{split}
A_{rs} &:= \frac{1}{2} (a_{rs} + \im b_{rs} - \im c_{rs} + d_{rs} ), \qq
B_{rs}:= \frac{1}{2} (a_{rs} - \im b_{rs} - \im c_{rs} - d_{rs} ), \\
C_{rs} &:= \frac{1}{2} (a_{rs} + \im b_{rs} + \im c_{rs} - d_{rs} ), \qq
D_{rs} := \frac{1}{2} (a_{rs} - \im b_{rs} + \im c_{rs} + d_{rs} ).
\end{split}
\ee
The $SO(p,q)$ condition for $\Lambda$ is converted into the condition
\be
M^{\mathrm{T}}(\Lambda) 
\mat{ 0 & \ & 1_N \\ 1_N & & 0 } M(\Lambda)
= \mat{ 0 & \ &  1_N \\ 1_N & & 0 }.
\ee
More explicitly,
$A^{\mathrm{T}} C + C^{\mathrm{T}} A = 0$,  
$B^{\mathrm{T}} D + D^{\mathrm{T}} B = 0$, 
$A^{\mathrm{T}} D + C^{\mathrm{T}} B = 1_N$.

First, let us discuss the transformation properties
of the local coordinate
 $(\gamma, u_{rs})$ for the pure spinor $\lambda$.
A pure spinor transforms as follows
\be
| \lambda' \rangle = S(\Lambda) | \lambda \rangle,
\qq
S(\Lambda) \Gamma^{a} n_{a}^{(I)} S^{-1}(\Lambda)
= \Gamma^{a} {n'}_{a}^{(I)}, \qq
S(\Lambda) = \exp\left( \frac{1}{4} \theta_{ab} \Gamma^{ab} \right).
\ee

If $\mathrm{det} ( C u + D) \neq 0$,
then we have
\be
\left(
A_r^{(+)} + A_s^{(-)} u'_{sr}
\right) | \lambda' \rangle = 0,
\qq
u' = ( A u + B)( C u + D)^{-1}.
\ee
So, the parameters $u = U T^{-1}$ transform fractionally
under the Lorentz transformation.
The Lorentz transformed pure spinor can be written as
\be
| \lambda' \rangle
= \gamma' \exp\left[ \frac{1}{2} u'_{rs} A_r^{(-)} A_s^{(-)}
\right] | + \rangle.
\ee

To summarize, under the Lorentz transformation $\Lambda$,
the parameters $(\gamma, u_{rs})$ transforms as follows
\be
\begin{split}
\gamma \rightarrow \gamma' &= \gamma \mathcal{F}_+(\Lambda, u), \\
u \rightarrow u' &= ( Au + B)( Cu+D)^{-1}.
\end{split}
\ee
Here
\bel{calF}
\mathcal{F}_+(\Lambda,u):=\langle +| S(\Lambda) \ex^U | + \rangle.
\ee
The forms of $\mathcal{F}_+$ for special cases are given in Appendix B.

Next, let us consider the transformation properties for $\omega$:
\be
| \omega' \rangle = S(\Lambda) | \omega \rangle.
\ee
The pure spinor gauge \eqref{pureomega} is preserved 
under the Lorentz transformation:
\be
| \omega' \rangle_{\mathrm{p}} = S(\Lambda) | \omega \rangle_{\mathrm{p}}.
\ee
And parameters $(\rho, w^{rs})$ transform as follows:
\be
\begin{split}
\rho & \rightarrow \rho' = \rho \mathcal{F}_-(\Lambda, w), \\
w & \rightarrow w' = ( D w + C )( B w + A)^{-1}.
\end{split}
\ee
Here
\be
\mathcal{F}_-(\Lambda, w) =
\langle - | S(\Lambda) \ex^W | - \rangle
= \langle + | \ex^{-W} S^{-1}(\Lambda) | + \rangle.
\ee

The simple gauge \eqref{simple} is not preserved
under the Lorentz transformation. So, in this case,
we should consider the accompanying ``gauge transformation'' \eqref{GTr}:
\be
S(\Lambda) | \omega \rangle_{\mathrm{s}}
= | \omega' \rangle_{\mathrm{s}} + | \xi( \Lambda, \omega_{\mathrm{s}} ) 
\rangle.
\ee

Let us examine the case of the ``bosonic ghost'' gauge more explicitly.

Note that
\be
\overline{\partial} u' 
= \left[ ( Cu+D)^{\mathrm{T}} \right]^{-1} \overline{\partial} u
( Cu + D)^{-1}.
\ee
Let us introduce an $N \times N$ antisymmetric matrix $\mathcal{G}$
by
\be
\mathcal{G}^{rs}:=\frac{\partial}{\partial u_{rs}} \log \mathcal{F}_+
= \frac{ \langle + | S(\Lambda) A_r^{(-)} A_s^{(-)} \ex^U | + \rangle}
{\langle + | S(\Lambda) \ex^U | + \rangle}.
\ee
If the conjugate fields transform as follows:
\be
\begin{split}
\beta & \rightarrow 
\beta' = \beta \mathcal{F}_+^{-1}, \\
v & \rightarrow
v' = ( Cu+D) ( v - \beta \gamma \mathcal{G} ) ( Cu+D)^{\mathrm{T}},
\end{split}
\ee
then the kinetic part of the 
action is invariant under the Lorentz transformations:
\be
\beta' \overline{\partial} \gamma' 
- \frac{1}{2} \mathrm{tr}( v' \overline{\partial} u')
= \beta \overline{\partial} \gamma 
- \frac{1}{2} \mathrm{tr}( v \overline{\partial} u).
\ee

\subsection{Equations of motion for pure spinors}

In this subsection, we derive the equations
of motion for pure spinors.
For simplicity, we take the simple ``bosonic ghost'' gauge
\eqref{purelambda} and \eqref{omegaghost}.

Then, the equations of motions 
for local coordinates
are easily obtained as
\be
\begin{split}
\overline{\partial} \gamma 
&= - \gamma \langle + | X \ex^U | + \rangle, \\
\overline{\partial} \beta 
&= \beta \langle + | X \ex^U | + \rangle, \\
\overline{\partial} u_{rs}
&= u_{rs} \langle + | X \ex^U | + \rangle
- \langle + | A_s^{(+)} A_r^{(+)} X \ex^U | + \rangle, \\
\overline{\partial} v^{rs}
&= - v^{rs} \langle + | X \ex^U | + \rangle
+ \left( \beta \gamma - \frac{1}{2} v^{pq} u_{pq} \right)
\langle + | X \ex^U A_r^{(-)} A_s^{(-)} | + \rangle \\
& \qq -
\langle + | V X \ex^U A_r^{(-)} A_s^{(-)} | + \rangle.
\end{split}
\ee
Using these equations, we can show that
the equations of motion for pure spinors 
(in the simple bosonic ghost gauge)
are given by
\bel{eom1}
\left( \overline{\partial} + X \right) | \lambda \rangle
= \Delta(u) X | \lambda \rangle, 
\ee
\bel{eom2}
- \overline{\partial}
\left({}_{\mathrm{s.b.}} \langle \omega^{\mathrm{T}} \mathcal{C} | \right)
+ {}_{\mathrm{s.b.}} \langle \omega^{\mathrm{T}} \mathcal{C} |  X
= {}_{ \mathrm{s.b.}}\langle \omega^{\mathrm{T}}  \mathcal{C} | X \Delta(u),
\ee
where
\be
\Delta(u) = \ex^U \left( \sum_{k=2}^{[N/2]} \mathcal{P}^{(2k)} \right)
\ex^{-U}
= \ex^U \left( \frac{1}{2}( 1 + \overline{\Gamma})
- \mathcal{P}^{(0)} - \mathcal{P}^{(2)} \right) \ex^{-U}.
\ee
Here $\mathcal{P}^{(n)}$ is the projector 
into the sector with a $U(1)$
charge $(N/2)-n$:
\be
\mathcal{P}^{(n)}:= \frac{1}{n!}
A_{r_1}^{(-)} A_{r_2}^{(-)} \dotsm A_{r_n}^{(-)} | + \rangle
\langle + | A_{r_n}^{(+)} \dotsm A_{r_2}^{(+)} A_{r_1}^{(+)}.
\ee
Here the $U(1)$ charge is an eigenvalue of the 
operator $A^{(+)}_r A^{(-)}_r - (N/2)$. 

To derive the equations of motion for $\omega$, 
we have used the relation
\be
\mathcal{P}^{(n)} U = U \mathcal{P}^{(n-2)}.
\ee
We can easily see that $\Delta(u)$ is a projector:
$\Delta^2(u) = \Delta(u)$, and it has the following properties:
\be
\Delta(u) | \lambda \rangle = 0, \qq
\Delta(u) | \overline{\partial} \lambda \rangle =0, \qq
{}_{\mathrm{s.b.}} \langle \omega^{\mathrm{T}} \mathcal{C}| \Delta(u) = 0.
\ee

To derive \eqref{eom1} and \eqref{eom2}, we have used
no properties of the interaction term $X$. Now, let us use
the ansatz of $X$ \eqref{rX}.

For $k=2,3,\dotsc, [N/2]$, we can see that
\be
\langle + | A^{(+)}_{r_{2k}} \dotsm A_{r_1}^{(+)}
| + \rangle =0, \qq
\langle + | A^{(-)}_{r_1} \dotsm A^{(-)}_{r_{2k}} | + \rangle = 0,
\ee
\be
\langle + | A^{(+)}_{r_{2k}} \dotsm A_{r_1}^{(+)}
\ex^{-U} \Gamma^{ab} \ex^U | + \rangle = 0, \qq
\langle + | \ex^{-U} \Gamma^{ab} \ex^U
A^{(-)}_{r_{1}} \dotsm A_{r_{2k}}^{(-)}
| + \rangle = 0.
\ee
Using these relations, for the special types of interactions
$X$ \eqref{rX}, we have
\bel{delzero}
\Delta(u) X | \lambda \rangle =0, \qq
\langle \omega^{\mathrm{T}} \mathcal{C} | X \Delta(u) = 0.
\ee
These relations can be understood as follows:
in order to have non-zero
$P^{(2k)} \ex^{-U} X \ex^U | + \rangle$ or
$\langle + | \ex^{-U} X \ex^U P^{(2k)}$, the operator $\ex^{-U} X \ex^U$
should contain components which change
the $U(1)$-charge by $2k$.
But the (complex) Lorentz transformed operator
$\ex^{-U} X \ex^U$ changes the $U(1)$-charge
at most $2$. Therefore, additional terms vanish for the
restricted interaction \eqref{rX}.

The relations \eqref{delzero}
lead to the final form of the equations of motion
for pure spinors:
\be
( \overline{\partial} + X ) | \lambda \rangle = 0,
\ee
\be
- \overline{\partial}
\left( {}_{\mathrm{s.b.}} \langle \omega^{\mathrm{T}}
\mathcal{C} | \right)
+ {}_{\mathrm{s.b.}} \langle \omega^{\mathrm{T}} \mathcal{C} |
X = 0.
\ee
Therefore, the equations of motion in the bosonic ghost gauge
retain the forms even if the effect of the pure spinor constraints
are taken into account.

\section{Pure spinor action without gauge symmetry}

In this section, we consider the pure spinor action 
with general interaction $X$.
Without restriction \eqref{rX},
the Lagrangian loses the invariance under the
gauge transformations \eqref{GTr}. Since these gauge symmetries are
no longer able to use to keep the ``simple gauge'',
one natural way to restrict the pair field $\omega$
is to require that it is also a pure spinor.
Another natural way is to impose no restriction on $\omega$.

We first consider the case with no restriction on $\omega$:
\be
\mathcal{L} = \langle \omega^{\mathrm{T}} \mathcal{C} | 
( \overline{\partial} + X) | \lambda \rangle.
\ee
The equations of motion for pure spinor $\lambda$ is simply given by
\be
( \overline{\partial} + X ) | \lambda \rangle = 0.
\ee
Using the solution of the pure spinor constraints:
$|\lambda \rangle = \gamma \ex^{U} | + \rangle$,
we can show that the equations of motion for $\omega$ is given by
\be
\Bigl( - \overline{\partial} (\langle \omega^{\mathrm{T}} \mathcal{C} |)
+ \langle \omega^{\mathrm{T}} \mathcal{C} | X \Bigr)
\ex^U P^{(0)} \ex^{-U} = 0,
\ee
\be
\Bigl( - \overline{\partial} (\langle \omega^{\mathrm{T}} \mathcal{C} |)
+ \langle \omega^{\mathrm{T}} \mathcal{C} | X \Bigr)
\ex^U P^{(2)} \ex^{-U} = 0.
\ee
Due to the pure spinor constraints for $\lambda$,
particular components of $- \overline{\partial} \langle \omega^{\mathrm{T}}
\mathcal{C} | + \langle \omega^{\mathrm{T}} \mathcal{C} | $
are required to obey the equations of motion.

The rest of degrees of freedom are not dynamical
and play the role of the Lagrange multiplier.
Since $\Delta(u) \overline{\partial} | \lambda \rangle = 0$,
the equation $( \overline{\partial} + X ) | \lambda \rangle =0$
implies the restriction on $X$: $\Delta(u) X | \lambda \rangle = 0$.
A solution for this restriction is given by \eqref{rX}.
So, in this case, the action goes back to the one with the gauge symmetry.
 
Next, let us consider the case that $\omega$ is a pure spinor.
It is convenient to use the pure ``bosonic ghost'' parametrization
\eqref{pbg}. 
With this parametrization, 
the Lagrangian density for the pure spinor pair is given by
\be
\mathcal{L}
= \beta \overline{\partial} \gamma 
+ \frac{1}{2} v^{rs} \overline{\partial} u_{rs}
+ \beta \gamma \langle + | \ex^{-Y} \ex^{-U}
X \ex^U | + \rangle.
\ee
The equations of motion for parameters are given by
\be
\begin{split}
\overline{\partial} \gamma
&= - \gamma \langle + | ( 1 + Y ) \ex^{-Y} \ex^{-U} X \ex^U | + \rangle, \\
\overline{\partial} u_{rs}
&= \langle + | A_r^{(+)} A_s^{(+)} \ex^{-Y} \ex^{-U} X \ex^U | + \rangle, \\
\overline{\partial} \beta
&= \beta \langle + | (1+Y) \ex^{-Y} \ex^{-U} X \ex^U | + \rangle, \\
\overline{\partial} y^{rs}
&= \langle + | \ex^{-Y} \ex^{-U} X \ex^U A_r^{(-)} A_s^{(-)} | + \rangle
- \langle + | \ex^{-Y} \ex^{-U} A_r^{(-)} A_s^{(-)} X
\ex^U | + \rangle.
\end{split}
\ee
Using these relations, we can show that the equations of motion
for pure spinors are given by
\be
\bigl( \overline{\partial} + X \bigr) | \lambda \rangle
= \Delta(\lambda, \omega) X | \lambda \rangle,
\ee
\be
- \overline{\partial} \langle \omega^{\mathrm{T}} \mathcal{C} |
+ \langle \omega^{\mathrm{T}} \mathcal{C} | X
= \langle \omega^{\mathrm{T}} \mathcal{C} | X \Delta(\lambda, \omega),
\ee
where a projector $\Delta(\lambda, \omega)$ is given by
\be
\Delta( \lambda, \omega)
= \ex^U \ex^Y \left( \sum_{k=2}^{[N/2]} 
\mathcal{P}^{(2k)} \right) \ex^{-Y} \ex^{-U}.
\ee
So, due to the pure spinor constraints for $\lambda$ and for $\omega$,
the equations of motion contain additional terms with
the projector $\Delta(\lambda, \omega)$.

If we restrict the interaction $X$ to the form \eqref{rX},
the operator $\ex^{-Y} \ex^{-U} X \ex^U \ex^Y$
changes the $U(1)$-charge at most $2$, therefore,
the additional terms vanish.
   
\section{Discussion}

In this paper, we considered the Berkovits type action
for the pure spinors in even dimensions.

We showed that
for general interactions $X$, 
both of $\lambda$ and $\omega$ are pure spinors and
the equations of motions for pure spinors
are modified from the naive ones. 
The extra terms contain
a particular projector $\Delta(\lambda, \omega)$.

If the types of interactions are restricted 
to the ``ghost number'' $u(1)$
and the Lorentz $so(p,q)$ current couplings, 
the action has a large ``gauge symmetry''. 
In this case, at least in the pure and simple
 ``bosonic ghost'' gauges, the extra terms vanish and
the equations of motion for the pure spinors retain 
the naive form even if the pure spinor constraints are taken 
into account.


\vspace{1cm}

{\bf Acknowledgments}\\
The author would like to thank N. Berkovits for useful comments
and INFN for financial support.
This work is partially supported by the EU network EUCLID, 
no. HPRN-CT-2002-00325.


\appendix

\section{Review of Pure Spinor}

Let us consider the Clifford algebra $C(p,q)$ associated to 
$V=\mathbb{R}^{p,q}$. The generators of $C(p,q)$ are
$2^m \times 2^m$ Gamma matrices
\be
\{ \Gamma^a, \Gamma^b \} = 2 \eta^{ab}, \qq
\eta_{ab} = \mathrm{diag}(+,+,\dotsm, +, -, - ,\dotsm, -), \qq
a, b = 1,2,\dots, D.
\ee
Here $D=p+q$ and $m = [D/2]$. Further, we require
that $(\Gamma^a)^{\dag} = \Gamma_a$.

A spinor $|\lambda \rangle \in \mathbb{C}^{2^m}$ is said to
be \textit{pure} \cite{car,che}
if we can take $m$ linear independent
complex vectors $n^{(I)}$ ($I=1,2,\dotsc,m$) in $V^{\mathbb{C}}
= V \otimes \mathbb{C} = \mathbb{C}^D$, 
the complexification of the vector space $V$, 
such that
\bel{purecond}
\Gamma^a n^{(I)}_a | \lambda \rangle = 0, \qq I=1,2,\dots, m.
\ee
In order to hold \eqref{purecond}, the complex vectors should satisfy
the \textit{isotropic} (or \textit{null}) conditions:
\be
n^{(I) a} n^{(J)}_{a} = 0, \qq I,J = 1,2,\dots, m.
\ee
So, the subspace $W = \mathrm{span}\{ n^{(1)}, \dotsc, n^{(m)} \}
\subset \mathbb{C}^{D}$ is a \textit{maximal isotropic} subspace.
Note that the isotropic conditions can be rewritten as follows:
\bel{MIC}
\{ \Gamma^{a} n_{a}^{(I)}, \Gamma^{b} n_{b}^{(J)} \} = 0.
\ee

In even dimension $(D=2N)$,
the complexification $V^{\mathbb{C}}$ decomposes into 
two isotropic spaces
with maximal dimension : $V^\mathbb{C} = W \oplus \overline{W}$.
This fact allows one 
to solve the pure spinor conditions in terms
of $N$ pairs of fermionic creation and annihilation operators
\cite{bud-tra}:
\be
\{ A^{(\epsilon_1)}_r, A^{(\epsilon_2)}_s \}
= \delta_{\epsilon_1+\epsilon_2, 0} \delta_{rs}, \qq
\epsilon_1, \epsilon_2 = \pm 1,
\ee
where
\be
A^{(\pm)}_r:= \frac{1}{2}( \widehat{\Gamma}^{2r-1} 
\pm \im \widehat{\Gamma}^{2r} ), \qq 
( A^{(\pm)}_r )^{\dag} = A^{(\mp)}_r, \qq 
r=1,2,\dots, N.
\ee
Here $\Gamma^{a}= \omega_{a} \widehat{\Gamma}^{a}$ with
$(\omega_{a})^2 = \eta_{aa}$ (\text{no sum}).
(So, $(\widehat{\Gamma}^a)^{\dag} = \widehat{\Gamma}^a$
and $\{ \widehat{\Gamma}^a, \widehat{\Gamma}^b \} = 2\delta^{ab}$).

The pure spinor conditions \eqref{purecond} can be rewritten as follows:
\be
\Gamma^{a} n_{a}^{(I)} | \lambda \rangle
= ( A^{(+)}_r T_r{}^I + A^{(-)}_r U_r{}^I ) | \lambda \rangle =0 ,
\ee
where
$T_r{}^I = \widehat{n}^{(I)}_{2r-1} + \im
\widehat{n}^{(I)}_{2r}$, and 
$U_r{}^I = \widehat{n}^{(I)}_{2r-1} - \im
\widehat{n}^{(I)}_{2r}$ 
for $r=1,2,\dotsc,N$ and $I=1,2,\dotsc, N$.
Here
$\widehat{n}^{(I)}_a:= \omega_a n^{(I)}_a$ (no sum).

The maximal isotropic condition \eqref{MIC} becomes
\bel{MIC2}
T_r{}^I U_r{}^J + U_r{}^I T_r{}^J = 
( T^{\mathrm{T}} U + U^{\mathrm{T}} T)^{IJ} = 0.
\ee

If $\mathrm{det}( T_r{}^I ) \neq 0$, then the pure spinor conditions
can be rewritten as
\bel{cohpure}
( A^{(+)}_r + A^{(-)}_s u_{sr} ) | \lambda \rangle =0,
\ee
where $u_{sr} := ( U T^{-1})_{sr} = U_s{}^I ( T^{-1})_{Ir}$.
From \eqref{MIC2}, $u$ is antisymmetric: $u_{sr} = - u_{rs}$.

The equations \eqref{cohpure} are easily solved as 
\bel{SP}
| \lambda \rangle 
= \gamma
\exp\left[ \frac{1}{2} u_{rs} A_r^{(-)} A_s^{(-)} \right] | + \rangle
= \gamma \prod_{1 \leq r< s \leq N}
\left( 1 + u_{rs} A_r^{(-)} A_s^{(-)} \right) | + \rangle,
\ee
where $|+ \rangle $ is the Fock vacuum:
$A^{(+)}_r | + \rangle = 0$ ($r=1,2,\dotsc, N$).

In general, for $\epsilon=(\epsilon_1, \epsilon_2, \dotsc, 
\epsilon_N)$, $\epsilon_i=\pm 1$, we can rewrite the elements
$\Gamma^a n^{(I)}_a$ as follows:
\be
\Gamma^a n^{(I)}_a = A_r^{(+)} T_r{}^I + A_r^{(-)} U_r{}^I
= A_r^{(\epsilon_r)} T^{(\epsilon)}_r{}^I 
+ A_r^{(-\epsilon_r)} U^{(\epsilon)}_r{}^I,
\ee
where
\be
\mat{ U^{(\epsilon)} \\ T^{(\epsilon)} }
= \mat{ 1_N - I^{(\epsilon)} & \ & I^{(\epsilon)} \\ 
I^{(\epsilon)} & & 1_N - I^{(\epsilon)} }
\mat{ U \\ T }.
\ee
Here
\be
I^{(\epsilon)} 
= \frac{1}{2} \mathrm{diag}( 1-\epsilon_1, 1-\epsilon_2, 
\dotsc, 1-\epsilon_N ),
\qq
(I^{(\epsilon)})_{rs} = \delta_{\epsilon_r, -1} \delta_{rs}.
\ee
If $\mathrm{det}( T^{(\epsilon)}) \neq 0$, 
the pure spinor conditions can
be written as
\be
\left[ A_r^{(\epsilon_r)} + A_s^{(-\epsilon_s)} u_{sr}^{(\epsilon)} \right]
| \lambda \rangle = 0, \qq
u^{(\epsilon)} = U^{(\epsilon)} ( T^{(\epsilon)} )^{-1}.
\ee
Therefore, the pure spinors for $\mathrm{det}(T^{(\epsilon)}) \neq 0$
can be expressed as
\be
| \lambda \rangle =\gamma^{(\epsilon)} 
\exp\left[ \frac{1}{2} u_{rs}^{(\epsilon)} A_r^{(-\epsilon_r)} 
A_s^{(-\epsilon_s)} \right] | \epsilon \rangle,
\ee
where
\be
| \epsilon \rangle:= 
\left( A_1^{(-)} \right)^{(1/2)(1-\epsilon_1)}
\left( A_2^{(-)} \right)^{(1/2)(1-\epsilon_2)} \dotsm 
\left( A_N^{(-)} \right)^{(1/2)(1-\epsilon_N)} | + \rangle.
\ee
The space of pure spinors are covered by $2^N$ open sets and
the local coordinates $(\gamma^{(\epsilon)}, u_{rs}^{(\epsilon)})$
parametrize pure spinors ``near'' $| \epsilon \rangle$.
Obvious examples of pure spinors which cannot be expressed
in the form \eqref{SP} are $|\epsilon \rangle$
for $\epsilon \neq (+1,+1,\dotsc, +1)$.
So, we need to consider these $2^N$ local coordinate systems
in order to express all pure spinors.

Note that the pure spinor has definite chirality:
\be
\overline{\Gamma} | \lambda \rangle = 
\left( \prod_{r=1}^N \epsilon_r \right) | \lambda \rangle.
\ee
Here the chirality matrix is chosen such that 
$\overline{\Gamma} | + \rangle = | + \rangle$.
So, there are $2^{N-1}$ choices of $\epsilon = ( \epsilon_1, \dotsc,
\epsilon_N)$ for each chirality.

For other properties of pure spinors, see, for example,  
\cite{LM,GP,PR,ber-che}.

\section{Explicit form of $\mathcal{F}_+(\Lambda, u)$
for special case}

For a Lorentz transformation in the $ab$-plane:
\be
S(\Lambda^{(ab)} ) = \exp\left( (\theta/2)
\widehat{\Gamma}^{ab} \right),
\ee
the function $\mathcal{F}_+(\Lambda,u)$ \eqref{calF} is given by
\be
\begin{split}
\mathcal{F}_+( \Lambda^{(2r-1,2r)}, u)
&= \ex^{\im (\theta/2)}, \\
\mathcal{F}_+( \Lambda^{(2r-1,2s-1)}, u)
&= \cos(\theta/2) - u_{rs} \sin (\theta/2), \qq r\neq s, \\
\mathcal{F}_+( \Lambda^{(2r-1,2s)}, u)
= \mathcal{F}_+( \Lambda^{(2r,2s-1)}, u)
&= \cos(\theta/2) + \im  u_{rs} \sin (\theta/2), \qq r\neq s, \\
\mathcal{F}_+( \Lambda^{(2r,2s)}, u)
&= \cos(\theta/2) + u_{rs} \sin (\theta/2), \qq r\neq s.
\end{split}
\ee

\end{document}